\documentclass[twocolumn,showpacs,preprintnumbers,amsmath,amssymb,floatfix,aps]{revtex4}

\usepackage{graphicx}
\usepackage{dcolumn}
\usepackage{bm}

\newcommand{\x}{{\mathbf{x}}}
\newcommand{\xp}{{\mathbf{x'}}}
\newcommand{\xs}{{\mathbf{x''}}}
\newcommand{\rv}{{\mathbf{r}}}

\begin{document}

\title{Released momentum distribution of a Fermi gas in the BCS-BEC
  crossover}

\author{M.L. Chiofalo$^{1}$, S. Giorgini$^{2,3}$ and M. Holland$^{2}$}
\address{$^{1}$INFM and Classe di Scienze, Scuola Normale Superiore,
  Pisa, Italy\\
  $^{2}$JILA and Department of Physics, University of Colorado at
  Boulder,
  Boulder, CO 80309-0440, U.S.A\\
  $^{3}$Dipartimento di Fisica, Universit\`a di Trento and BEC-INFM,
  I-38050 Povo, Italy}

\date{\today}

\begin{abstract}
  We develop a time-dependent mean-field theory to investigate the
  released momentum distribution and the released energy of an
  ultracold Fermi gas in the BCS-BEC crossover after the scattering
  length has been set to zero by a fast magnetic-field ramp. For a
  homogeneous gas we analyze the non-equilibrium dynamics of the
  system as a function of the interaction strength and of the ramp
  speed. For a trapped gas the theoretical predictions are compared
  with experimental results.
\end{abstract}


\maketitle
The study of the momentum distribution of an atomic gas in the quantum
degenerate regime carries a wealth of information on the role played
by interactions and on the existence of a superfluid order parameter.
As an example, in a homogeneous Bose gas at $T=0$, the momentum
distribution $n_{\bf k}$ exhibits a singular behavior at small wave
vectors $n_{\bf k}\simeq mc/2\hbar k$, which is determined by the
excitation of phonons propagating with the speed of sound $c$ and is a
signature of Bose-Einstein condensation~\cite{PS}. In a corresponding
system of fermions with attractive interactions, the broadening of the
Fermi surface is instead a consequence of the formation of pairs and
of the presence of a superfluid gap~\cite{Gennes}. This latter effect
becomes dramatic in the BCS-BEC crossover region where the pairing gap
is of the order of the Fermi energy of the system~\cite{BCS-BEC}. The
second moment of the momentum distribution defines the kinetic energy
of the system: $E_{kin}=\sum_{\bf k}n_{\bf k}\hbar^2k^2/(2m)$, where
$m$ is the mass of the atoms. This quantity, which also plays a
central role in the many-body description of ultracold gases, is very
sensitive to the large-$k$ behavior of $n_{\bf k}$. For interacting
systems the dominant contribution to $E_{kin}$ comes from short-range
correlations, where the details of the interatomic potential are
relevant. In the case of a zero-range potential it is well known that
the momentum distribution decreases like $1/k^4$ for large momenta and
the kinetic energy diverges in dimensionality greater than one. This
unphysical divergence can be understood recalling that the zero-range
approximation is only correct to describe the region of momenta $k\ll
1/r_0$, where $r_0$ denotes the physical range of
interactions~\cite{Nikitin}. This behavior of the kinetic energy is a
general feature of quantum-degenerate gases, where interactions are
well described by the $s$-wave scattering length $a$, holding both for
fermions and bosons and for repulsive and attractive
interactions~\cite{note1}.

The physics of ultracold gases is characterized by a clear separation
of energy scales: the energy scale associated with the two-body
physics as fixed for example by $\hbar^2/mr_0^2\sim 10 mK$, being
$r_0\sim 100 a_0$ the typical interaction length of the Van der Waals
potential, and the energy scale associated with the many-body physics
as determined by the typical Fermi energy $\epsilon_F\sim 1 \mu K$.
This separation of energy scales provides a very large range of
timescales for which the dynamical process can be safely considered
{\it fast} ({\it diabatic}) as the many-body dynamics is concerned and
{\it slow} ({\it adiabatic}) with respect to the two-body dynamics.
This feature is exploited in recent experiments aiming to measure the
momentum distribution, that are based on the ballistic expansion of
the cloud after the scattering length has been quickly set to zero by
a fast magnetic-field ramp~\cite{Bourdel,Regal}. These experiments
give access to the released momentum distribution, which is a
non-equilibrium quantity defined as the momentum distribution of the
system after the scattering length has been rapidly ramped to $a=0$.
Provided the timescale of the ramp satisfies the conditions given
above, the released momentum distribution does not depend on the
detailed structure of the interatomic potential, being in this sense
universal, but it does depend on the timescale of the ramping process.
An important theoretical issue is to investigate the properties of the
released momentum distribution and to understand what useful
information about the system can be extracted from it.

In this Letter we calculate the released momentum distribution of a
Fermi gas at $T=0$ in the BCS-BEC crossover.  We calculate the
released momentum distribution and its second moment for a
homogenoeous system as a function of the interaction strength
$1/(k_Fa)$, where $k_F$ is the Fermi wave vector. For harmonically
trapped systems we give an explicit prediction of the column
integrated released momentum distribution and of the released energy
for values of the interaction strength ranging from the BCS to the BEC
regime and we compare our results with recently obtained experimental
data~\cite{Regal}.

We consider an unpolarized two-component Fermi gas with equal
populations of the $\uparrow$ and $\downarrow$ components:
$N_\uparrow=N_\downarrow=N/2$, where $N$ is the total number of
particles. We determine the dynamical evolution of such a system
starting from the equations of motion for the non-equilibrium density
matrices of the $\uparrow$ and $\downarrow$ components interacting
through the Hamiltonian
\begin{eqnarray}
H&=&\sum_\sigma \int d\x\,
\psi^\dagger_\sigma(\x)\left(-\frac{\hbar^2\nabla^2_\x}{2m}\right)
\psi^{}_\sigma(\x)\\
&+&\int
d\x\, d\xp \psi^\dagger_\uparrow(\x)\psi^\dagger_\downarrow(\xp)V(\x ,\xp)
\psi^{}_\downarrow(\xp)\psi^{}_\uparrow(\x)\nonumber
\label{eqH}
\end{eqnarray}
where $\sigma=\uparrow,\downarrow$ labels the spins and $V(x,x')$ is
the interaction potential to be specified later. The correlations are
treated within a mean-field approach~\cite{BCS-BEC}, where they are
expressed in terms of the normal $G_{N\sigma}(\x ,\xp ,t) =
\bigl<\psi^\dagger_\sigma(\xp ,t)\psi_\sigma(\x ,t)\bigr>$ and the
anomalous $G_{A}(\x ,\xp ,t) = \bigl<\psi_{\downarrow}(\xp
,t)\psi_{\uparrow}(\x ,t)\bigr>$ density matrices. By neglecting the 
Hartree terms one obtains the following
coupled equations of motion~\cite{Gennes} (from now on
$G_{N\uparrow}(\x ,\xp ,t) = G_{N\downarrow}(\x ,\xp ,t)\equiv G_N(\x
,\xp ,t)$)
\begin{eqnarray}
  i\hbar\frac{dG_N(\x ,\xp ,t)}{dt}
  =\left(-\frac{\hbar^2\nabla^2_\x}{2m}+\frac{\hbar^2\nabla^2_\xp}{2m}\right)
  G_N(\x ,\xp ,t)\nonumber\\
  +\int d\xs [V(\x ,\xs)-V(\xs ,\xp)]G_A(\x ,\xs ,t)G_A^\ast(\xs ,\xp ,t)
\label{eqGN}
\end{eqnarray}   
and
\begin{eqnarray}
i\hbar\frac{dG_A(\x ,\xp ,t)}{dt}
=\left(-\frac{\hbar^2\nabla^2_\x}{2m}-\frac{\hbar^2\nabla^2_\xp}{2m}\right)G_A(\x ,\xp ,t)\nonumber\\
+V(\x ,\xp)G_A(\x ,\xp ,t)\nonumber\\
- \int d\xs V(\x ,\xs)G_N(\xp ,\xs
  ,t)G_A(\x ,\xs ,t)\nonumber\\
-\int d\xs V(\x ,\xs)G_N(\x ,\xs ,t)G_A(\xs ,\xp ,t)\; .
\label{eqGA}
\end{eqnarray}   
The short-range nature of the interaction potential $V(\x ,\xp)$ can
be properly described through the regularized pseudopotential
$V(\rv)=(4\pi a\hbar^2/m)\delta(\rv)(\partial/\partial
r)r$~\cite{Castin}, with $a$ the $s$-wave scattering length and
$r\equiv|\x-\xp|$. During the magnetic-field ramp the value of the
scattering length changes in time according to the relation
\begin{equation}
a(t)=a_{bg}\left(1-\frac{\Gamma}{B(t)-B_0}\right) \;,
\label{resonance}
\end{equation}
valid close to the Feshbach resonance. In the above expression,
$a_{bg}$ denotes the background scattering length, $B_0$ and $\Gamma$
the position and width of the resonance respectively, and $B(t)$ is
the instantenous value of the magnetic field. Under the dynamical
conditions that we are considering, where the non-equilibrium
processes take place over a timescale adiabatic with respect to the
two-particle problem and diabatic with respect to the many-particle
system, the time evolution does not depend on the details of the
short-range potential and the effect of interactions results in a
boundary condition at short length scales
\begin{equation}
\left[\frac{(rG_A(r,t))'}{rG_A(r,t)}\right]_{r=0}=-\frac{1}{a(t)}\; ,
\label{eqBC}
\end{equation}
where the prime indicates the derivative with respect to $r$, which
must be fulfilled at any time $t$.  For small values of $r$, many-body
effects in Eq. (\ref{eqGA}) can be neglected and the boundary
condition (\ref{eqBC}) corresponds to the one of the two-body problem
with the pseudopotential $V(\rv)$, where $G_A(r,t)$ plays the role of
the wave function for the relative motion. A similar argument is known
to hold also in the electron gas, where one has the boundary condition
$[g'(r)/g(r)]_{r=0}=1/a_B$ for the pair correlation function $g(r)$
($a_B$ is the Bohr radius), referred to as the Kimball
relation~\cite{Kimball}.

In the case of a homogeneous system and by using the pseudopotential
approximation for the interatomic potential $V(\x , \xs)$ and the
boundary condition (\ref{eqBC}), Eqs. (\ref{eqGN})-(\ref{eqGA}) can be
greatly simplified.  For example, the term $\int d\xs V(\x
,\xs)G_N(\xp ,\xs ,t)G_A(\x ,\xs ,t)$ in Eq.~(\ref{eqGA}) becomes
$(4\pi\hbar^2a/m)G_N(r ,t)[(rG_A)']_{r=0}=-(4\pi\hbar^2/m)G_N(r
,t)[(rG_A)]_{r=0}$, where we have also used the fact that the short
ranged $V(\x ,\xs)$ picks up the $\xs=\x$ value of $G_N(\xp ,\xs ,t)$.
After dealing with the other terms in a similar way, we obtain the
simpler coupled equations for $\tilde{G}_N(r,t)\equiv rG_N(r,t)$ and
$\tilde{G}_A(r,t)\equiv rG_A(r,t)$
\begin{eqnarray}
i\hbar\frac{d\tilde{G}_N(r,t)}{dt}&=&\frac{8\pi\hbar^2}{m}i\Im 
\left( \tilde{G}_A(r,t)[\tilde{G}_A^\ast(t)]_{r=0}
\right) 
\label{eqCoup1}\\
i\hbar\frac{d\tilde{G}_A(r,t)}{dt}&=&-\frac{\hbar^2\partial^2}{m\partial
  r^2}\tilde{G}_A(r,t)\nonumber\\
&+&\frac{8\pi\hbar^2}{m}\tilde{G}_N(r,t)
[\tilde{G}_A(t)]_{r=0}\; ,
\label{eqCoup2}
\end{eqnarray}   
with the boundary condition
$[(\tilde{G}_A)'/\tilde{G}_A]_{r=0}=-1/a(t)$. Notice that interaction
effects only enter Eqs. (\ref{eqCoup1})-(\ref{eqCoup2}) through the
boundary condition (\ref{eqBC}). We determine the initial conditions
$\tilde{G}_N(r,t=0)$ and $\tilde{G}_A(r,t=0)$ of
Eqs.(\ref{eqCoup1})-(\ref{eqCoup2}) from the mean-field gap and number
equations corresponding to the equilibrium state of the gas with the
initial value of the scattering length $a(0)$
\begin{eqnarray}
  n&=&\int_0^\infty \frac{dk k^2}{2\pi^2}
  \left(1-\frac{\epsilon_k-\mu}{\sqrt{(\epsilon_k-\mu)^2+\Delta^2}}\right)
\label{eqEQ1}\\
\frac{m}{4\pi\hbar^2a(0)}&=&\int_0^\infty \frac{dk k^2}{4\pi^2}
\left[\frac{1}{\epsilon_k}
  -\frac{1}{\sqrt{(\epsilon_k-\mu)^2+\Delta^2}}\right]\; ,
\label{eqEQ2}
\end{eqnarray}
where $\epsilon_k=\hbar^2k^2/2m$, $\mu$ is the chemical potential,
$\Delta$ the superfluid gap and $n=n_\uparrow+n_\downarrow$ the total
particle density. The functions $\tilde{G}_N$ and $\tilde{G}_A$ are
then calculated from the Bogoliubov quasiparticle amplitudes
\begin{equation}
  u_k^2=1-v_k^2=\frac{1}{2}\left(1+\frac{\epsilon_k-\mu}
    {\sqrt{(\epsilon_k-\mu)^2+\Delta^2}}\right) \; ,
\label{equkvk}
\end{equation}      
as
\begin{equation}
\tilde{G}_N(r,t=0)=\frac{1}{2\pi^2}\int_0^\infty dk k\sin (kr) v_k^2
\label{eqGNin}
\end{equation}
and
\begin{equation}
  \tilde{G}_A(r,t=0)=\frac{1}{2\pi^2}\int_0^\infty dk k\sin (kr) u_kv_k \; .
\label{eqGAin}
\end{equation}
We solve the dynamic equations (\ref{eqCoup1})-(\ref{eqCoup2}) with
the initial conditions (\ref{eqGNin})-(\ref{eqGAin}) and $a(t)$ given
by (\ref{resonance}) from the initial time $t=0$ to the final time
$t=t_f$, where $a(t_f)=0$. The released momentum distribution is then
calculated from the Fourier transform of $G_N$ at the time $t=t_f$
\begin{equation}
  n_{\bf k}(t=t_f)=\int d\rv e^{i{\mathbf k}\cdot\rv} G_N(r,t=t_f) \; .
\label{eqnkrel}
\end{equation}

\begin{figure}
\begin{center}
\includegraphics*[width=7.0cm]{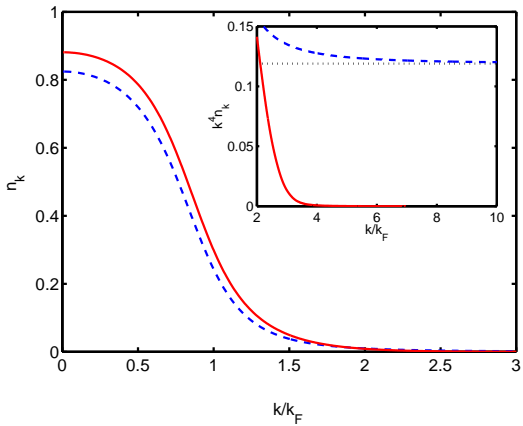}
\caption{(color online) Released momentum distribution (solid lines)
  of a homogeneous gas at unitarity, $1/(k_Fa(0))=0$, for a ramp rate
  of 2$\mu$s/G. The large-$k$ behavior of $n_{\bf k}$ weighted by
  $k^4$ is shown in the inset, where the dotted line corresponds to
  the equilibrium asymptotic value $(\Delta/2\epsilon_F)^2$.  The
  initial equilibrium distribution is also shown (dashed lines).}
\label{fig1}
\end{center}
\end{figure}

\begin{figure}
\begin{center}
\includegraphics*[width=7.0cm]{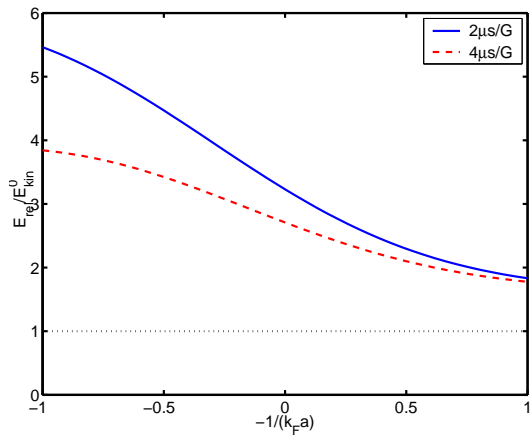}
\caption{(color online) Released energy as a function of the interaction strength for two values of the ramp 
rate. The energy is normalized to the kinetic energy of the non-interacting gas $E_{kin}^0=3\epsilon_F/8$.}
\label{fig2}
\end{center}
\end{figure}

The results for the homogeneous gas are shown in
Figs.~\ref{fig1}-\ref{fig2}. In Fig.~\ref{fig1} we compare the
equilibrium momentum distribution $n_{\bf k}(t=0)$ in the unitary
limit, $1/(k_Fa(0))=0$, with the corresponding released momentum
distribution (\ref{eqnkrel}) calculated for a magnetic-field ramp rate
of 2$\mu$s/G. For values of $k\lesssim k_F$ the shape of the
distribution does not change appreciably. The large-$k$ tail is
instead greatly suppressed (as is shown in the inset) and the second
moment of the released $n_{\bf k}$ is a convergent integral. Notice
that the fast decaying tail of the released $n_{\bf k}$ affects the
normalization constant.  In Fig.~\ref{fig2} we show the results of the
released energy as a function of the initial interaction strength
$1/(k_Fa(0))$ for two different values of the magnetic-field ramp
rate.  We notice that on the BCS side of the crossover, $k_Fa(0)<0$,
the dependence on the ramp rate is weak, while on the BEC side,
$k_Fa(0)>0$, a faster ramp produces a significantly larger energy. In
the BEC regime the system is in fact more sensitive to changes of the
high-energy tail of the momentum distribution. Deep in the BCS regime,
$-1/(k_Fa(0))\gg 1$, the released energy reduces to the kinetic energy
of the non-interacting gas $E_{kin}^0=3\epsilon_F/8$. In the opposite
BEC regime, $-1/(k_Fa(0))\ll-1$, many-body effects become less
relevant and the released energy coincides with the one obtained from
the dissociation of the molecular state~\cite{Regal} (see
Fig.~\ref{fig4}).

In order to make quantitative comparison with the experiment, we now
consider harmonically trapped systems confined by the external
potential $V_{ext}({\bf r})=m(\omega_x^2x^2+
\omega_y^2y^2+\omega_z^2z^2)/2$. Within the local density
approximation (LDA) we introduce the rescaled spatial variables
$\tilde{x}=x\sqrt{\omega_x/\omega}$,
$\tilde{y}=y\sqrt{\omega_y/\omega}$ and $\tilde{z}=x
\sqrt{\omega_z/\omega}$, so that the confining potential becomes
isotropic in the new coordinates $V_{ext}({\bf r})=
m\omega^2\tilde{R}^2/2$, where
$\omega=(\omega_x\omega_y\omega_z)^{1/3}$ is the geometric average of
the harmonic oscillator frequencies. For each spatial slice
$\tilde{{\bf R}}=(\tilde{\x}+\tilde{\x}^\prime)/2$, Eqs.
(\ref{eqEQ1})-(\ref{eqEQ2}) are solved for the local chemical
potential $\mu_{local}(\tilde{{\bf R}})$ and the local density
$n(\tilde{{\bf R}})$, subject to the normalization $\int
d^3\tilde{{\bf R}}n(\tilde{{\bf R}})=N$ and the local equilibrium
condition $\mu=\mu_{local}(\tilde{{\bf R}})+V_{ext}(\tilde{{\bf R}})$.
Each slice is then evolved according to Eqs.
(\ref{eqCoup1})-(\ref{eqCoup2}) with initial conditions
$\tilde{G}_{N}(\tilde{r},\tilde{R},t=0)$ and
$\tilde{G}_{A}(\tilde{r},\tilde{R},t=0)$, where $\tilde{{\bf
    r}}=\tilde{\x}-\tilde{\x}^\prime$ is the relative coordinate. The
released momentum distribution is obtained from
$G_{N}(\tilde{r},\tilde{R},t)$ at the final time $t=t_f$ through the
integral over the rescaled coordinates $\tilde{{\bf R}}$ and
$\tilde{{\bf r}}$,
\begin{equation}
  n(k,t=t_f)=\int d^3\tilde{{\bf R}} \int d^3\tilde{{\bf r}} 
  e^{i{\mathbf k}\cdot\tilde{{\bf r}}} 
  G_N(r,R,t=t_f) \;.
\label{eqMD}
\end{equation}

\begin{figure}
\begin{center}
\includegraphics*[width=7.0cm]{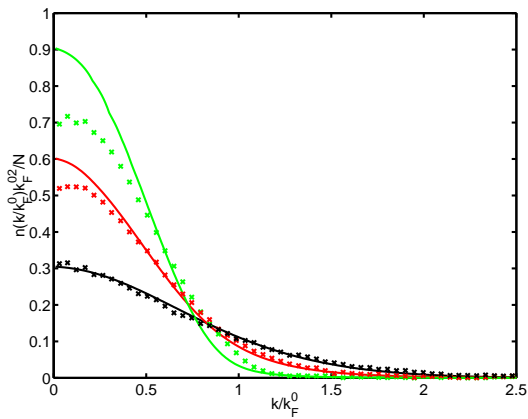}
\caption{(color online) Column integrated released momentum
  distribution of a harmonically trapped gas. From top to bottom, 
  the lines correspond to $1/(k_F^0a(0))=-0.66$ (green), 
  $1/(k_F^0a(0))=0$ (red) and $1/(k_F^0a(0))=0.59$ (black). The 
  magnetic-field ramp rate is 2$\mu$s/G. The symbols correspond to 
  the experimental results of Ref.~\cite{Regal}.}
\label{fig3}
\end{center}
\end{figure}

\begin{figure}
\begin{center}
\includegraphics*[width=7.0cm]{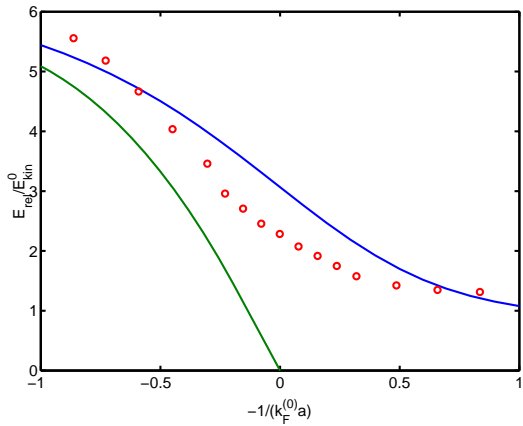}
\caption{(color online) Released energy of a harmonically trapped gas
  as a function of the interaction strength $1/(k_F^0a(0))$ for a ramp
  rate of 2$\mu$s/G (upper blu line). The lower (green) line is the 
  corresponding result solving the two-body problem associated with the 
  molecular state. The symbols are the experimental results from
  Ref.~\cite{Regal}. The energy is normalized to the kinetic energy of
  the non-interacting gas $E_{kin}^0=3\epsilon_F^0/8$.}
\label{fig4}
\end{center}
\end{figure}

In Fig.~\ref{fig3} we compare the column integrated released momentum
distribution $n(\sqrt{k_x^2+k_y^2},t_f)= \int_{-\infty}^\infty dk_z
n(k,t_f)$, calculated from Eq. (\ref{eqMD}), with the experimental
results obtained in Ref. \cite{Regal}. The values of the interaction
strength are $1/(k_F^0a(0))=$-0.66, 0 and 0.59 as in the experiments
and the magnetic-field ramp rate is 2$\mu$s/G. The agreement is quite
good for the momentum distribution on the BEC side of the resonance
and in the unitary limit. On the BCS side of the resonance the
experimental $n(k)$ is more broadened because the Hartree mean-field
term, which enhances the shrinking of the cloud due to attraction, is
neglected in the calculation. In Fig.~\ref{fig4} we show the results
for the released energy of the inhomogeneous gas as a function of the
interaction strength $1/(k_F^0a(0))$, where
$k_F^0=(24N)^{1/6}\sqrt{m\omega/\hbar}$ is the Fermi wave vector in
the center of the trap corresponding to a non-interacting gas.
Experimental results from Ref.~\cite{Regal} and theoretical results
obtained by solving the time-dependent Schr\"odinger equation for the
molecular state (see \cite{Regal}) are also shown in Fig.~\ref{fig4}.
The full many-body calculation reduces to the molecular two-body
result only in the deep BEC regime, $-1/k_F^0a(0)\ll -1$, and agrees
better with the experimental results. Considering that there are no
adjustable parameters in the comparison between theory and experiment,
the agreement is remarkable over the whole crossover region. In the
unitary limit the mean-field approach employed in the present study is
known to overestimate the equilibrium energy per particle compared to
more advanced quantum Monte Carlo calculations \cite{MC}. This might
be the reason for the larger energy obtained around resonance compared
to the observed one. Furthermore, on the BCS side of the resonance,
the present approach neglects the mean-field Hartree term and we
expect a faster convergence to the kinetic energy of the non
interacting gas $E_{kin}^0=3\epsilon_F^0/8$, where
$\epsilon_F^0=(\hbar k_F^0)^2/2m$, as $-1/k_F^0a(0)$ becomes large. A
more sophisticated time-dependent theory is needed to improve the
quantitative agreement with experiments.

In conclusion, we have developed a time-dependent mean-field scheme
which allows one to calculate the released momentum distribution and
the released energy of a Fermi gas in the BCS-BEC crossover if the
scattering length is set to zero using a fast magnetic-field ramp. Our
analysis clarified that the dynamical effects due to the
magnetic-field ramp suppress the high-energy tail of the momentum
distribution. For harmonically trapped systems we compared our
theoretical predictions with the recently obtained experimental
results of Ref.~\cite{Regal}. Qualitatively we reproduced the data
well, but significant quantitative discrepancies are found. The
underestimate of the release energy of the theory compared to
experiment on the deep BCS side is anticipated due to the neglection
of the Hartree energies. On the deep BEC side, the underestimate may
be due to the effect of finite temperature on the experimental data in
that regime. The significant overestimate of the release energy at
resonance is more difficult to explain and may indicate the inadequacy
of the Leggett ground state form for describing the gas in the unitary
limit. This would illustrate a need to include correlations beyond the
pair correlations given in the Cooper pair picture in order to explain
the observed experimental expansion energetics on resonance.

Acknowledgements: SG acknowledges support by the Ministero
dell'Istruzione, dell'Universit\`a e della Ricerca (MIUR) and from the
National Science Foundation. MH acknowledges support from the U.S.
Department of Energy, Office of Basic Energy Sciences via the Chemical
Sciences, Geosciences and Biosciences Division.

\end{document}